\documentclass[aps,showpacs,twocolumn,amsmath,amssymb,floatfix,
nofootinbib,preprintnumbers]
{revtex4}
\usepackage{bm}
\usepackage{epsfig}

\begin{document}

\title{Above-Threshold Poles in Model-Independent Form Factor
Parametrizations}

\author{Benjam\'{i}n Grinstein}
\email{bgrinstein@ucsd.edu}
\affiliation{Department of Physics, University of California, San Diego,
La Jolla, California 92093, USA}

\author{Richard F. Lebed}
\email{richard.lebed@asu.edu}
\affiliation{Department of Physics, Arizona State University, Tempe,
Arizona 85287-1504, USA}

\date{September, 2015}

\begin{abstract}
  The model-independent parametrization for exclusive hadronic form
  factors commonly used for semileptonic decays is generalized to
  allow for the inclusion of above-threshold resonant poles of known
  mass and width.  We discuss the interpretation of such poles,
  particularly with respect to the analytic structure of the relevant
  two-point Green's function in which they reside.  Their presence has
  a remarkably small effect on the parametrization, as we show
  explicitly for the case of $D \to \pi e^+ \nu_e$.
\end{abstract}



\maketitle


\section{Introduction} \label{sec:Intro}

Studies of the analytic structure of Green's functions in quantum
field theory (QFT) have a long and illustrious history.  Here we
merely outline, in the briefest possible description, one specific
line of inquiry on two-point Green's functions, ultimately stretching
back to QFT fundamentals like the optical theorem and the
K\"{a}ll\'{e}n--Lehmann spectral decomposition, and ending with a
practical yet rigorous parametrization for the form factors of
semileptonic decays of hadrons, in which a heavy quark flavor
($s,c,b$) decays to a lighter one.

The starting point is the two-point Green's function of two currents,
in our case a conjugate pair $J J^\dagger$ of weak-interaction
currents $J^\mu \equiv \bar q \Gamma^\mu Q$, where $\Gamma^\mu$ is the
$V \! - \! A$ weak interaction Lorentz structure (at least at leading
perturbative order) responsible for changing the heavy quark flavor
$Q$\@.  This two-point function is analytic everywhere in the plane of
complex momentum $q^2$, except at poles corresponding to resonances
and cuts corresponding to collections of particles going onto the mass
shell.  The most important one is the so-called {\it unitarity cut\/}
corresponding to the production of the lightest pair of hadrons (of
flavor content $Q\bar q$ plus its conjugate) from the currents, since
it has the lowest branch point on the real $q^2$ axis.  In 1963,
Meiman~\cite{Meiman:1963} was the first to consider the conformal
mapping of the entire cut $q^2$ plane to the unit disk in a variable
$z$, and to note the existence of a bound on the coefficients of the
powers of $z$ for any function derived from the two-point function.
Some years later,
Okubo~\cite{Okubo:1971jf,Okubo:1971my,Okubo:1972ih,Shih:1972qn}
applied the $z$-variable transformation to the two-point function
relevant to the semileptonic process $K_{\ell 3}$, to obtain bounds on
certain moments of the form factors.  In 1980, Bourrely {\it et
al.}~\cite{Bourrely:1980gp} showed how to obtain bounds for the form
factors by using the evaluation of the two-point function in a region
where perturbative QCD is applicable.  Finally, in the mid-1990s, Boyd
{\it et al.}~\cite{Boyd:1994tt,Boyd:1995cf,
Boyd:1995sq,Boyd:1995tg,Boyd:1997qw,Boyd:1997kz} showed how
below-threshold poles---essential to properly treating the analytic
structure of the two-point function---can be incorporated into the
$z$-expansion by means of a well-known trick of complex analysis
called {\it Blaschke factors} (the analytic significance of which for
heavy-hadron form factors was first noted by
Caprini~\cite{Caprini:1994fh,Caprini:1994np}), and applied the
$z$-parametrization thus derived to a number of heavy-quark
semileptonic decays.

It is then natural to ask whether the {\em above}-threshold poles, for
which the corresponding resonances can decay to on-shell pairs of
mesons with the quantum numbers of $JJ^\dagger$, can loosen or perhaps
even fatally weaken these bounds and the parametrization following
from them.  In fact, as to be shown here, even the most extreme case
of a prominent resonance just barely above threshold does not
significantly damage the quality of the parametrization.  As a {\em
  formal\/} matter, only poles on the first Riemann sheet in $q^2$
change the analytic structure of the form factor, while the (optional)
inclusion of poles on other sheets (in particular, above-threshold
resonances) changes only the unitarity bound, and therefore does {\em
  not\/} affect the analytic structure.  Nevertheless, we argue in
this paper that the maximum effect of such an above-threshold pole on
the unitarity bound is such that it can be treated as though it arises
from a first-sheet pole at the corresponding $q^2$ location.  In order
to support this conclusion rigorously, one must develop a technology
in which the above-threshold poles can be treated solely according to
their analytic structure within the two-point function.

The contribution of this paper is to show that above-threshold poles,
corresponding to resonances of known mass and width, can also be
accommodated into the parametrization by using Blaschke factors.  The
essential mathematical point is that, by virtue of possessing a finite
width, the poles lie off the unitarity cut and therefore can be
treated as if they reside inside the unit circle, where Blaschke
factors are applicable.  An important subtlety to be discussed below
is the sense in which resonant poles, which first appear on the second
Riemann sheet, can be handled in this way.  The essential
phenomenological point is that, by virtue of the widths being
sufficiently small compared to the resonant mass (which in turn lies
above the heavy-quark hadronic threshold), the poles lie {\em
barely\/} inside the unit circle, and the corresponding corrections
from the Blaschke factors weaken the bounds on coefficients of the
semileptonic form factor parametrization very little.  As a specific
example, one might expect the $D^*$ resonance, which lies very close
to the $D \pi$ threshold, to have a pronounced effect on $D^{+,0} \to
\pi^{0,-} \ell^+ \nu_\ell$ form factors, but we show below that the
effect is only at the level of 1 part in $10^{-3}$.  The loosening of
the bounds actually becomes more prominent for lighter quarks; but
even in the case of $K_{\ell 3}$, the $K^*$ pole is seen only to
loosen the bounds by a few percent.

This paper is organized as follows: In Sec.~\ref{sec:review} we review
the current technology of the $z$-expansion.
Section~\ref{sec:newmath} generalizes the expansion to the case of a
pole lying above the pair-production threshold of the two-point
function but slightly off the unitarity cut due to a finite imaginary
part.  In Sec.~\ref{sec:exist} we address the question of what sense
in which an observed resonance can be treated as such a pole
contributing to the dispersion relation from which the expansion is
derived.  Section~\ref{sec:Example} shows the effect of this approach
in two sample cases, the decays $D^{+,0} \to \pi^{-,0} \ell^+
\nu_\ell$ and $K^{+,0} \to \pi^{0,-} \ell^+ \nu_\ell$, and in
Sec.~\ref{sec:Concl} we offer concluding remarks.

\section{Review of the Expansion} \label{sec:review}

We reprise here the key formulae relevant to the form factor
parametrization in a description almost identical to that in
Ref.~\cite{Boyd:1997kz}, and incorporate minor modifications relevant
to the inclusion of above-threshold resonant poles.  Starting with the
heavy-light ($Q \to q$) vectorlike ($V$, $A$, or $V \! - \! A$) quark
transition current
\begin{equation} \label{eq:Jdefn}
J^\mu \equiv \bar Q \Gamma^\mu q \, ,
\end{equation}
the two-point momentum-space Green's function $\Pi_J^{\mu \nu}$ is
defined, and separated into manifestly spin-1 ($\Pi_J^T$) and spin-0
($\Pi_J^L$) pieces, by
\begin{eqnarray}
\Pi_J^{\mu\nu} (q) & \equiv & i \! \int \! d^4 x \, e^{iqx} \left< 0
\left| T J^\mu (x) J^{\dagger \nu} (0) \right| 0 \right> \nonumber \\ 
& = & \frac{1}{q^2} \left( q^\mu q^\nu - q^2
  g^{\mu\nu} \right) \Pi^T_J (q^2) + \frac{q^\mu q^\nu}{q^2} \Pi^L_J
(q^2) \, . \nonumber \\
\label{eq:twopoint}
\end{eqnarray}
In QCD, the functions $\Pi^{L,T}_J$ contain divergences of different
degrees and must undergo subtractions (one and two, respectively) to
appear in finite dispersion relations:
\begin{eqnarray}
\chi^L_J (q^2) \equiv \frac{\partial \Pi^L_J}{\partial q^2} & = &
\frac{1}{\pi} \int_0^\infty \! dt \, \frac{{\rm Im} \,
  \Pi^L_J(t)}{(t-q^2)^2} \, , \nonumber \\
\chi^T_J (q^2) \equiv \frac 1 2 \frac{\partial^2 \Pi^T_J}{\partial
  (q^2)^2} & = & \frac{1}{\pi} \int_0^\infty \! dt \, \frac{{\rm Im}
  \, \Pi^T_J(t)}{(t-q^2)^3} \, . \label{eq:chis}
\end{eqnarray}
Perturbative QCD (or more thoroughly, QCD sum rules) may be used to
compute the functions $\chi (q^2)$ at values of $q^2$ far from the
region where $J$ can produce manifestly nonperturbative effects like
resonances.  This condition specifically reads $(m_Q + m_q)
\Lambda_{\rm QCD} \ll (m_Q + m_q)^2 - q^2$.  $q^2 = 0$ is sufficient
for $Q = c, b$, while $Q = s$ might require a slightly negative value,
say $q^2 = -1$~GeV$^2$.

The functions ${\rm Im} \, \Pi_J$ are evaluated by inserting into the
dispersion relation a complete set of states $X$ that couple the
current $J$ to the vacuum, leading to
\begin{equation} \label{eq:unitary}
{\rm Im} \, \Pi^{T,L}_J (q^2) = \frac 1 2 \sum_X (2\pi)^4 \delta^4
(q - p_X) \left| \left< 0 \left| J \right| \! X \right> \right|^2 \, .
\end{equation}
This relation shares a common origin with the optical theorem and the
K\"{a}ll\'{e}n--Lehmann spectral decomposition, but it refers
particularly to matrix elements of a specific current $J$ (in our
case, the amplitudes for the weak processes $W^* \to X$) rather than
those of a single field or a full transition operator.  The dispersion
relations Eqs.~(\ref{eq:chis}) indicate the equality of the
perturbatively evaluated function $\chi (q^2)$ with an integral over
the production rate as a function in momentum of the processes $W^*
\to X$, which includes phase space and other smooth functions.  Since
each term in the sum is semipositive definite, one obtains a strict
inequality for each $X$, which may further be restricted if one
chooses to include other states in the sum.  In our case, we choose
$X$ to be the two-particle states consisting of the lightest meson
pair in which one of them contains a $Q$ quark (mass $M$) and the
other a $\bar q$ (mass $m$).  For $D_{\ell 3}$ ($K_{\ell 3}$) decays,
$X$ is $D \pi$ ($K \pi$).  Defining
\begin{equation} \label{eq:tpmdef}
t_\pm \equiv (M \pm m)^2 \, ,
\end{equation}
and choosing, for definiteness, the form factor $F(t)$ to be the one
coupling to $\Pi^T$, one has
\begin{equation} \label{eq:FFbound1}
\frac{1}{\pi \chi^T (q^2)} \int_{t_+}^\infty \! dt \frac{W(t) \,
|F(t)|^2}{(t-q^2)^3} \leq 1 \, ,
\end{equation}
where $W(t)$ is a simple, computable nonnegative function (largely
phase space factors).  An analogous expression holds for $\Pi^L$.

The complex-$t$ plane contains a branch cut extending from $t_+ \to
\infty$.  It is mapped to the unit disk in a variable $z$ (with the
two sides of the cut forming the unit circle $C$) using the conformal
variable transformation
\begin{equation} \label{eq:zdef}
z(t;t_0) \equiv \frac{\sqrt{t_+ - t} - \sqrt{t_+ - t_0}} {\sqrt{t_+ -
    t} + \sqrt{t_+ - t_0}} = \frac{t_0 - t} {\left( \sqrt{t_+ -
    t} + \! \sqrt{t_+ - t_0} \right)^2} ,
\end{equation}
where $t_0$ is a parameter chosen later for convenience.  In
particular, $z$ is real for $t \le t_+$ and a pure phase for $t \ge
t_+$.  In fact, the definition in Eq.~(\ref{eq:zdef}) can be used in
several capacities since, as seen from its second form, multiplying by
$z(t;t_0)$ eliminates a simple pole $t = t_0$.  The bound
Eq.~(\ref{eq:FFbound1}) on $F(t)$ may then be rewritten as
\begin{equation} \label{eq:FFbound2}
\frac{1}{\pi} \int_{t_+}^\infty \! dt \left| \frac{dz(t;t_0)}{dt}
\right| \left| \phi (t;t_0) P(t) F(t) \right|^2 \leq 1 \, ,
\end{equation}
where the weight function $\phi(t;t_0)$ is called an {\it outer
function\/} in complex analysis.  It is given here by
\begin{equation} \label{eq:outer}
\phi(t;t_0) = \tilde P(t) \left[ \frac{W(t)}{|dz(t;t_0)/dt| \, \chi^T
    (q^2) (t-q^2)^3} \right]^{1/2} \, ,
\end{equation}
where the function $\tilde P(t)$ is a product of factors $z(t;t_s)$ or
$\sqrt{z(t;t_s)}$ (and hence unimodular on the unit circle $|z(t;t_0)|
= 1$) designed to remove {\em kinematical\/} singularities at points
$t = t_s$ from the other factors in Eq.~(\ref{eq:outer}).  The
functions $\phi(t;t_0)$ for any form factor of spin-0 and spin-1 meson
and spin-$\frac 1 2$ baryon semileptonic decays are tabulated in
Ref.~\cite{Boyd:1997kz}.  On the other hand, the function $P(t)$ in
Eq.~(\ref{eq:FFbound2}) is a product of Blaschke factors $z(t;t_p)$
(again unimodular on the unit circle $|z(t;t_0)| = 1$) that remove
{\em dynamical\/} singularities due to resonant poles in the two-point
function.

In total, the analyticity of the two-point function away from the cut
and all poles is most efficiently expressed by isolating the factors
that encode the nonanalytic behavior of the form factor $F(t)$ into
the functions $\phi(t;t_0)$ and $P(t)$ and then transforming to the
variable $z = z(t;t_0)$, so that the dispersion relation inequality
Eq.~(\ref{eq:FFbound1}) or (\ref{eq:FFbound2}) becomes
\begin{equation} \label{eq:FFrelnz}
\frac{1}{2\pi i} \oint_C \frac{dz}{z} | \phi(z) P(z) F(z) |^2 \le 1 \,
,
\end{equation}
which in turn allows the expansion
\begin{equation} \label{eq:param}
F(t) = \frac{1}{|P(t)| \phi(t;t_0)} \sum_{n=0}^\infty a_n z(t;t_0)^n
\, ,
\end{equation}
with the bound of Eq.~(\ref{eq:FFrelnz}) now reading
\begin{equation} \label{eq:coeffs}
\sum_{n=0}^\infty a_n^2 \leq 1 \, .
\end{equation}
All possible functional dependences of the form factor $F(t)$
consistent with Eqs.~(\ref{eq:chis}) are now incorporated into the
coefficients $a_n$ of Eq.~(\ref{eq:param}), which are highly
constrained by Eq.~(\ref{eq:coeffs}).

The strength of the parametrization Eq.~(\ref{eq:param}) becomes truly
apparent when one notes that the kinematical variable $z$ typically
assumes a small range for semileptonic decays, so that the series
converges quickly and can be truncated after a small number of terms.
To be specific, let us rewrite Eq.~(\ref{eq:zdef}) in terms of parent
and daughter velocity 4-vectors $v^\mu \equiv p_M^\mu / M$, $v'^\mu
\equiv p_m^\mu / m$.  A convenient commonly used invariant is their
dot product,
\begin{equation} \label{eq:wdef}
w \equiv v \cdot v' = \gamma_m = \frac{E_m}{m} =
\frac{M^2 + m^2 - t}{2Mm} \, ,
\end{equation}
where $\gamma_m$ is the relativistic dilation factor of the daughter
$m$ in the rest frame of the parent $M$.  In terms of $w$,
Eq.~(\ref{eq:zdef}) becomes
\begin{equation} \label{eq:zinw}
z(t;t_0) = z(w;N) = \frac{\sqrt{1+w} - \sqrt{2N}} {\sqrt{1+w} +
  \sqrt{2N}} \, ,
\end{equation}
where $N$ is a free parameter related to $t_0$ by
\begin{equation} \label{eq:Ndefn}
N = \frac{t_+ - t_0}{t_+ - t_-} \, .
\end{equation}

The kinematic limits for the semileptonic decay $M \to m \ell \nu_\ell$
are $t_{\rm min} = m_\ell^2$, $t_{\rm max} = t_-$, which correspond,
respectively, to
\begin{eqnarray}
  w_{\rm max} & = & \frac{1 + r^2 - \delta^2}{2r} \, , \nonumber \\
  w_{\rm min} & = & 1 \, , \label{eq:wlimits}
\end{eqnarray}
or
\begin{eqnarray}
  z_{\rm max} & = & \frac{\sqrt{(1+r)^2 - \delta^2} - 2\sqrt{Nr}}
  {\sqrt{(1+r)^2 - \delta^2} + 2\sqrt{Nr}} \, , \nonumber \\
  z_{\rm min} & = & - \left( \frac{\sqrt{N} - 1} {\sqrt{N} + 1}
  \right) \, , \label{eq:zlimits}
\end{eqnarray}
using the abbreviations $r \equiv m/M$, $\delta \equiv m_\ell /M$.
The minimum (optimized) truncation error is achieved when $z_{\rm min}
= -z_{\rm max}$, which occurs when
\begin{equation} \label{eq:Nopt}
N_{\rm opt} = \sqrt{\frac{(1+r)^2 -\delta^2}{4r}} \, ,
\end{equation}
or
\begin{equation} \label{eq:t0opt}
  t_0 = t_+ \left[ 1 - \sqrt{ \left( 1- \frac{t_-}{t_+} \right)
    \left( 1 - \frac{m_\ell^2}{t_+} \right) } \right] \, .
\end{equation}

Evaluating at $N = N_{\rm opt}$, one finds
\begin{equation} \label{eq:zmaxminopt}
  z_{\rm max} = -z_{\rm min} = \frac{ \left[ (1+r)^2 - \delta^2
    \right]^{1/4} - (4r)^{1/4} }{ \left[ (1+r)^2 - \delta^2
    \right]^{1/4} + (4r)^{1/4} } \, ,
\end{equation}

While the Blaschke factors due to resonant poles at $t = t_p$ can be
expressed as $z(t;t_p)$, it is more convenient to use the form used in
previous works:
\begin{equation} \label{eq:Blaschke_old}
P(z;z_p) = \frac{z-z_p}{1-z z_p} \, ,
\end{equation}
and $z(t;t_p) = P(z;z_p)$ whenever $t_p < t_+$ (a subthreshold pole)
so that $z_p$ is real.  However, the same technique works just as well
for any complex value for $z_p$ inside the unit disk.  In that case,
the definition of Eq.~(\ref{eq:Blaschke_old}) can be generalized to
\begin{equation} \label{eq:Blaschke_new}
P(z;z_p) = \frac{\left| z_p \right|}{z_p} \frac{z_p - z}{1 - z_p^* z}
\, ,
\end{equation}
which, assuming $t_0 < t_+$, equals $z(t;t_p)$ times the phase of $t_p
- t_0$, the latter factor being irrelevant in the bound
Eq.~(\ref{eq:FFrelnz}).  Note that $P(0;z_p) = |z_p|$ ({\it i.e.},
with this definition $P(0;z_p)$ is manifestly nonnegative), and that
all $z_p$ with $|z_p| = 1$ give $P(z) = 1$.  The usefulness of the
Blaschke factors for phenomenology is determined by how much they
degrade the bound Eq.~(\ref{eq:param}) in the semileptonic region
(near $z=0$): Fewer poles with $|z_p| < 1$ means a more constrained
allowed region for $F(z)$.

\section{Poles Above Threshold} \label{sec:newmath}

Consider a pole at the complex mass value $M_R - i\Gamma/2$, such that
$M_R \equiv M + m + \Delta m > M+m = \sqrt{t_+}$ and $\Gamma > 0$.
Specifically, let us define dimensionless mass excess and width
parameters:
\begin{eqnarray}
  \mu & \equiv & \frac{\Delta m}{\sqrt{t_+}} = \frac{M_R}{\sqrt{t_+}}
  - 1 \, , \label{eq:mudef} \\
  \gamma & \equiv & \frac{\Gamma}{2\sqrt{t_+}} \, . \label{eq:gamdef}
\end{eqnarray}
It is furthermore advantageous to define the following dimensionless
variables:
\begin{eqnarray}
a     & \equiv & \mu (2 + \mu) -\gamma^2 \, , \nonumber \\
b     & \equiv & 2\gamma (1 + \mu) \, , \nonumber \\
c     & \equiv & \sqrt{ a^2 + b^2 } = \sqrt{ (\mu^2 + \gamma^2) [( 2 +
  \mu)^2 + \gamma^2 ]} \, , \nonumber \\
\beta & \equiv & \sqrt{1 - \frac{t_+}{t_0}} = \frac{2\sqrt{Nr}}{1+r}
\, . \label{eq:polevars}
\end{eqnarray}
One expects both $\mu \ll 1$, indicating that the mass does not lie
far above threshold, and $\gamma \ll 1$, indicating a narrow width.
The usual narrow-width approximation, $\Gamma \ll M_R$, can be
enhanced in this case to assume that the width is sufficiently small
so as to clearly separate the peak from threshold, $\gamma \ll \mu$
($\Gamma \ll \Delta m$).  Likewise, one expects $b \ll a \simeq c \ll
1$, but generically $\beta = O(1)$.  The specific values for the case
of $D^0 \to \pi^- e^+ \nu_e$, for which the $D^{*+}$ pole lies
slightly above the $D^0 \pi^+$ threshold, are presented in
Table~\ref{tab:Dnums}.\footnote{We use central values from
\cite{Agashe:2014kda}.  Strictly speaking, only the hadronic part of
$\Gamma_{D^{*+}}$ should be included in these strong-interaction
dispersion relations; however, the hadronic branching fraction of
$D^{*+}$ is $98.4 \pm 0.7 \%$, and therefore is taken to be 1.}.
Similar values hold for $D^+ \to \pi^0 e^+ \nu_e$ and for muon
channels.

\begin{table}[ht]
  \caption{Parameter values for the decay $D^0 \to \pi^- e^+ \nu_e$.}
  \label{tab:Dnums}
\begin{tabular}{cl|cl}
  $r$ & $7.484 \cdot 10^{-2}$ & $\mu$ & $2.919 \cdot 10^{-3}$ \\
  $\delta$ & $2.740 \cdot 10^{-4}$ & $\gamma$ & $2.080 \cdot 10^{-5}$
  \\
  $N_{\rm opt}$ & $1.964$ & $a \simeq c$ & $5.846 \cdot 10^{-3}$ \\
  $z_{\rm max} = -z_{\rm min}$ & 0.1672 & $b$ & $4.173 \cdot 10^{-5}$
  \\
  $\beta$ & $0.7135$ & $z_p$ & $-0.9762 + 0.2117 i$ \\
  & & $|z_p|$ & $0.99924$ \\
  & & $\arg z_p$ & $167.8^\circ$
\end{tabular}
\end{table}

Regardless of the smallness of any parameters, one can compute
compact closed-form solutions for the position of $z_p$.  One finds
\begin{equation} \label{eq:zp}
z_p = \frac{-\beta^2 + c + i \beta \sqrt{2(c+a)}} {\beta^2 + c + \beta
\sqrt{2c(c-a)}} \, ,
\end{equation}
from which one obtains
\begin{equation} \label{eq:abszp}
  \left| z_p \right|^2 = 1 - \frac{2\beta \sqrt{2(c-a)}} {\beta^2 +
    c + \beta \sqrt{2(c-a)}} \, ,
\end{equation}
and
\begin{equation} \label{eq:argzp1}
\arg z_p = \arg \left[ \left( -\beta^2 + c \right) + i \beta
\sqrt{2(c+a)} \right] \, .
\end{equation}
Using Eq.~(\ref{eq:argzp1}) with $\beta^2 > c$ (a resonance near
threshold), one has
\begin{equation} \label{eq:argzp2}
  \arg z_p = \pi - \tan^{-1} \left( \frac{\beta \sqrt{2(c+a)}}
    {\beta^2 - c} \right) \, ,
\end{equation}
while for $\beta^2 < c$ (a resonance far above threshold),
\begin{equation} \label{eq:argzp3}
\arg z_p = \tan^{-1} \left( \frac{\beta \sqrt{2(c+a)}} {c-\beta^2}
\right) \, .
\end{equation}
Neglecting $m_\ell$ ($\delta$), using $N = N_{\rm opt}$ from
Eq.~(\ref{eq:Nopt}), and retaining only the lowest power in $\Gamma$
($\gamma$), one obtains
\begin{equation} \label{eq:abszapprox}
1 - |z_p| \to \frac{\Gamma}{2\sqrt{\Delta m} (Mm)^{1/4}} \cdot
\frac{\beta^2}{\beta^2 + \mu (2 + \mu)} \cdot \frac{1+\mu}
{\sqrt{1+\mu/2}} \, ,
\end{equation}
while the argument of the arctangent in
Eqs.~(\ref{eq:argzp2})--(\ref{eq:argzp3}) becomes
\begin{equation} \label{eq:arctanarg}
\frac{2\sqrt{\Delta m}}{(Mm)^{1/4}} \cdot \frac{\beta^2
\sqrt{1+\mu/2}} {\left|\beta^2 - \mu (2+\mu) \right|} \, ,
\end{equation}
independent of the width to linear order.  Additionally taking the
near-threshold resonance limit $\mu \ll 1$, the latter two factors of
Eq.~(\ref{eq:abszapprox}) and the second factor of
Eq.~(\ref{eq:arctanarg}) become unity:
\begin{equation} \label{eq:abszpapprox2}
1 - |z_p| \to \frac{\Gamma}{2\sqrt{\Delta m} (Mm)^{1/4}} \, ,
\end{equation}
and
\begin{equation} \label{eq:argzpapprox}
\arg z_p \to \pi - \arctan \left[ \frac{2\sqrt{\Delta m}} {(Mm)^{1/4}}
  \right] \, .
\end{equation}
The corresponding exact values of $z_p$, $|z_p|$, and $\arg z_p$ for
$D^0 \to \pi^- e^+ \nu_e$ also appear in Table~\ref{tab:Dnums}.  The
values obtained from the approximate forms in
Eqs.~(\ref{eq:abszpapprox2}) and (\ref{eq:argzpapprox}) agree with
the exact results of Eqs.~(\ref{eq:abszp}) and (\ref{eq:argzp1})
within $10^{-5}$ and $0.15^\circ$, respectively.

The naive effect of such an additional pole is to allow $|F(z)|$---and
hence each of the coefficients $a_n$ in
Eqs.~(\ref{eq:param})--(\ref{eq:coeffs})---to be larger by a factor of
$1/|P(z;z_p)|$, where $z \in [-z_{\rm max} , z_{\rm max}]$ for the
semileptonic decay.  Noting that $|z_p|$ lies very close to
unity---much closer to unity than it does to the allowed semileptonic
values of $z$---one finds $1/|P(z;z_p)|$ to lie uniformly close to
unity, meaning that {\em the presence of a pole with $|z_p| \simeq 1$
weakens the model-independent form factor bounds very little}.  To
give a simple figure of merit, consider the value of $1/|P(z;z_p)|$ at
the center of the semileptonic range, $z=0$; as we have seen,
$1/|P(0;z_p)| = 1/|z_p|$.  The exact value is given by
Eq.~(\ref{eq:abszp}):
\begin{equation} \label{eq:Pexact}
\frac{1}{|P(0;z_p)|} = \frac{1}{|z_p|} = \left[ 1 - \frac{2\beta
\sqrt{2(c-a)}} {\beta^2 + \beta \sqrt{2(c-a)} + c} \right]^{-1/2} \, ,
\end{equation}
while its approximate value ($\mu \ll 1$) is given by
Eq.~(\ref{eq:abszpapprox2}):
\begin{equation} \label{eq:Papprox}
\frac{1}{|P(0;z_p)|} \to  1 + \frac{\Gamma}{2\sqrt{\Delta m}
(Mm)^{1/4}} \, .
\end{equation}
Again, inasmuch as the pole represents a resonance with a narrow width
well separated from threshold, $\Gamma \ll \Delta m \ll M + m$, the
correction term is quite small; in the case of $D^0 \to \pi^- e^+
\nu_e$, the allowed ranges for the $a_n$ are expanded by less than 8
parts in $10^4$.  In summary, the parametrization is exactly as before
in Eq.~(\ref{eq:param}), but the allowed range for each $a_n$ is
slightly expanded beyond $|a_n| \le 1$.

Of course, $z=0$ is just one point in the allowed range for
semileptonic decay.  Since the poles of interest lie not far above
threshold, $z_p$ lies rather close to $-1$; therefore, from
Eq.~(\ref{eq:Blaschke_new}), the largest correction to the $a_n$
factors occurs at $z = -z_{\rm max}$ ($t = m_{\ell}^2$).  In the case
of $D^0 \to \pi^- e^+ \nu_e$, the correction is still only about 1
part in $10^{-3}$.  The effect of the near-threshold pole is truly
minimal.

\section{Existence and Nature of Above-Threshold Poles} \label{sec:exist}

In the previous section, we have shown that incorporating an
above-threshold pole into the two-point function that corresponds to a
resonance is mathematically not difficult.  Here we discuss in detail
issues related to the question of whether such a treatment is
appropriate to physical resonances.

The most common approach treats an above-threshold resonance, which is
identified by a Lorentzian distribution in energy identified with a
Breit-Wigner distribution:
\begin{equation} \label{eq:Breit}
| {\cal M} |^2 \propto \frac{1}{(s - M_R^2)^2 + s \Gamma^2} \, ,
\end{equation}
as being associated with a Breit-Wigner pole at the value $\sqrt{s} =
M_R -i \Gamma/2$, assuming the narrow-width approximation $\Gamma \ll
M_R$.\footnote{Here we use $q^2 = s$ rather than $t$, to emphasize the
pair-production origin of the cut.}  More generally, the width
$\Gamma$ need not be a constant but can have an energy dependence,
$\Gamma(s)$.  In either case, one anticipates the existence of a pole
in the amplitude ${\cal M}$ off the real-$s$ axis.

Nevertheless, as was pointed out long ago~\cite{Schwinger:1960}, an
observable lineshape arbitrarily close to an idealized Breit-Wigner
distribution can be simulated even in the absence of a literal pole
off the real-$s$ axis.  Inasmuch as most complex energy values are
experimentally inaccessible, the only ways to unambiguously detect a
literal pole (either measuring at the pole location itself or
measuring at points surrounding it and using Cauchy's theorem) are
unavailable.  So while the presence of a pole in the complex plane is
a natural way to interpret the appearance of a narrowly peaked
distribution along the real axis, its certainty is not
guaranteed~\cite{Paz:2015}.  One may model the amplitude along the cut
by incorporating an explicit Lorentzian function, including a specific
value of residue~\cite{Bhattacharya:2011ah,Epstein:2014zua}.  See
also~\cite{Ananthanarayan:2011uc}, in which the resonance is
incorporated into phase and modulus information along the cut.

Even so, the assumption of a pole at a complex value of $z$ near the
unit circle has been seen in the previous section to loosen the bounds
on semileptonic form factors very little.  Note particularly that the
Blaschke factor Eq.~(\ref{eq:Blaschke_new}) makes reference only to
the position of the pole and not its residue; therefore, it must work
equally well for a pole with residue as large as is allowed by
unitarity (which is explicitly built into the dispersion relation) and
a pole with vanishing residue---which is, of course, no pole at all.
Since, once again, the effect of a complex-valued pole projected along
the real axis is to allow for a narrow peak of an arbitrary physically
allowed value of residue, one sees that including the Blaschke factor
in the two-point function is appropriate for accommodating the effects
of a Breit-Wigner lineshape along the real-axis cut, but does not
actually commit one to demanding the existence of a pole off the real
axis.

Another interesting point regarding the above-thresh\-old pole is its
appearance in the full Riemann surface for the two-point function.
The existence of a cut indicates the existence of at least one
additional Riemann sheet.  For example, a particle pair created in the
$L^{\rm th}$ partial wave has phase space proportional to $k^{2L+1}$,
where
\begin{equation} \label{eq:kdefn}
k = \sqrt{\frac{(s-t_+)(s-t_-)}{4s}} \,
\end{equation}
is the center-of-momentum-frame value of the spatial momentum of
either particle.  Since the discontinuity along the cut is
proportional to phase space, one thus obtains a two-sheet Riemann
surface, corresponding to the double valuedness of the square root
function.  The number of sheets doubles each time a distinct two-body
threshold is encountered.

The question then becomes, on what sheet do the physical resonances
live, and on what sheet or sheets were the dispersion relation
integrals obtained?  The first question was originally answered by
Peierls~\cite{Peierls:1954}, who argued that a resonant pole must live
on the unphysical (second) Riemann sheet below the real axis, just on
the other side of the cut from the first sheet.  Otherwise, the
Schwarz reflection principle would require a pole just below the real
axis on the first sheet to have a mirror pole just above the real axis
on the first sheet; and since the negative imaginary value
($-i\Gamma/2$) of the former pole is necessary to obtain an
exponentially decaying state, the latter mirror pole would correspond
to an unphysical runaway state.

On the other hand, the contour bounding the dispersion integral is
easily seen to live entirely on the first sheet, since its derivation
uses the Schwarz reflection principle to obtain a nonnegative
contribution along the cut.  So then, one may ask, why worry about
poles that are not even encircled by the contour?  The answer is
simple pragmatism: A pole that lies just below the cut on the second
sheet creates a Breit-Wigner projection along the cut identical to the
contribution that would be obtained from an unphysical pole just above
the cut on the first sheet.  While causality knows that the pole lies
just below the cut on the second sheet, the dispersion relation is
sensitive to the pole only through its projection along the cut, and
this contribution can be obtained from a pole at $M_R \pm i\Gamma/2$
on any sheet such that its projection along the real axis agrees with
data.  One sees that treating the pole as if it occurred in the fourth
quadrant of the first sheet, as done in Sec.~\ref{sec:newmath}, leads
to the appropriate projection along the cut.  In particular, the value
of $\arg z_p$ given in Eq.~(\ref{eq:argzp2}) places it in the second
quadrant of the complex-$z$ plane, but the value of $\arg z_p$
symmetric about $\pi$ lying in the third quadrant of the complex-$z$
plane is equally valid for the analysis of Sec.~\ref{sec:newmath}.

This point is worth emphasizing.  The physical second-sheet poles do
{\em not\/} literally appear inside the unit circle $|z| = 1$.  The
first-sheet poles examined in Sec.~\ref{sec:newmath} are strictly
unphysical.  However, such an unphysical pole near the unitarity cut,
were it nevertheless to occur, would create a Breit-Wigner lineshape
indistinguishable from that created by a physical second-sheet pole
equally near the unitarity cut.  The unphysical poles of
Sec.~\ref{sec:newmath} must not be thought of as altering the {\em
analytic structure\/} of the form factor---in other words, of changing
the shape of the form factor $F(z)$ through the $z$-dependence of the
Blaschke factor $P(z)$.  Rather, they alter the {\em unitarity
bound\/} Eq.~(\ref{eq:coeffs}), by allowing the coefficients $a_n$ to
have larger ranges in exchange for the benefit of completely ignoring
the effect of an above-threshold resonance, no matter how prominent.

It is interesting to note that the leading-order perturbative
expansion of the two-point function $\chi(q^2)$ in the deep Euclidean
region contains logarithmic dependence (and polylogarithmic dependence
at higher perturbative order).  As is well known, these functions have
Riemann surfaces with an infinite number of sheets, in contrast to the
two sheets for a function with a half-integer power, such as those
previously discussed.  Since the perturbative two-point function can
be considered as an inclusive sum over all allowed exclusive channels,
the mismatch between the sheet counting can be construed as indicating
the necessity of including an arbitrarily large number of open
channels in order to achieve quark-hadron duality.

\section{Examples} \label{sec:Example}

\subsection{$D$ Semileptonic Decays}

The unflavored semileptonic decays of $D$ mesons are particularly
interesting for this formalism.  First, several such modes ($D \to \{
\pi , \, \rho , \, \omega , \, \eta , \, \eta^\prime \}$) have been
observed, each with an $O(10^{-3})$ branching fraction.  Furthermore,
modes with both $e^+$ and $\mu^+$ have been seen.  Since all of these
modes proceed through the $J^\mu = \bar q \Gamma^\mu c$ currents,
where $q$ is a light quark and $\Gamma^\mu$ represents Lorentz
structure, they all serve to saturate the same small set of dispersion
relations, leading to stronger bounds on any one of them.

Second, the processes $D^{+,0} \to \pi^{0,-} \ell^+ \nu_\ell$ are
remarkable due to the closeness of the $D^*$ resonance to the
crossed-channel $D \pi$ threshold in each case.
Specifically~\cite{Agashe:2014kda},
\begin{eqnarray}
m_{D^{*0}} - m_{D^0} - m_{\pi^+} & = & 5.86 \pm 0.07 \ {\rm MeV} \, ,
  \nonumber \\
m_{D^{*+}} - m_{D^+} - m_{\pi^0} & = & 5.68 \pm 0.08 \ {\rm MeV} \, .
\end{eqnarray}
As we have seen, the smallness of these numbers (combined with the
small width $\Gamma_{D^{*+}} = 83.4 \pm 1.8$~keV) guarantees a minimal
modification to the allowed range for the semileptonic form factor
coefficients $a_n$.  Furthermore, isospin symmetry relates the two
processes\footnote{The $D^{*0}$ width has only a measured upper bound
of 2.1~MeV~\cite{Agashe:2014kda}, but isospin symmetry predicts it to
be close to that of $D^{*+}$.} (separately for the $I= \frac 1 2$ and
$\frac 3 2$ channels, but with no resonance in the latter channel).
As seen in Ref.~\cite{Boyd:1997kz}, the presence of separate
isospin-related channels increases the function $\phi(z)$ in
Eq.~(\ref{eq:outer}) by a Clebsch-Gordan factor $\sqrt{n_I}$, where
$n_I = \frac 3 2$ for $D \to \pi$.  Noting that $\phi(z)$ appears in
the denominator of the parametrization Eq.~(\ref{eq:param}), one finds
that the coefficient bound of Eq.~(\ref{eq:coeffs}) effectively has
its unity factor replaced by $\frac 2 3$---a much more dramatic effect
than that due to the near-threshold $D^*$ pole.

\subsection{$K$ Semileptonic Decays}

The $K_{\ell 3}$ decays are interesting in this context, part\-ly
because they were the ones originally studied by
Okubo~\cite{Okubo:1971jf,Okubo:1971my,Okubo:1972ih,Shih:1972qn}, but
also because they possess a prominent, fairly narrow resonance $K^*$
($M_R = 891.66$~MeV, $\Gamma = 50.8$~MeV) that lies significantly far
above the threshold $\sqrt{t_+} = m_K + m_\pi$.  It is worth pointing
out that the $K_{\ell 3}$ and $D_{\ell 3}$ decays have the same form
factor and isospin structure.  For definiteness, let us consider the
specific mode $K^+ \to \pi^0 e^+ \nu_e$, for which the numerical
values of the key parameters are presented in Table~\ref{tab:Knums},
but the corresponding values for the modes $K_L \to \pi^- e^+ \nu_e$,
$K^+ \to \pi^0 \mu^+ \nu_\mu$, and $K_L \to \pi^- \mu^+ \nu_\mu$ are
very similar.  It should also be noted that the decay $\tau \to K
\pi \nu_\tau$ is bounded by the same dispersion relation, and indeed,
can provide a tighter constraint~\cite{Hill:2006bq}.

\begin{table}[ht]
  \caption{Parameter values for the decay $K^+ \to \pi^0 e^+ \nu_e$.}
  \label{tab:Knums}
\begin{tabular}{cl|cl}
  $r$ & $0.2734$ & $\mu$ & $0.4184$ \\
  $\delta$ & $1.035 \cdot 10^{-3}$ & $\gamma$ & $4.040 \cdot 10^{-2}$
  \\
  $N_{\rm opt}$ & $1.218$ & $a$, $c$ & $1.010$, $1.017$ \\
  $z_{\rm max} = -z_{\rm min}$ & $4.919 \cdot 10^{-2}$ & $b$ &
  $0.1146$ \\
  $\beta$ & $0.9062$ & $z_p$ & $0.1006 + 0.9400 i$ \\
  & & $|z_p|$ & $0.94535$ \\
  & & $\arg z_p$ & $83.9^\circ$
\end{tabular}
\end{table}

The large distance of the resonant mass from threshold is manifested
in the angle of $z_p$ lying much further from $\pi$ radians, indeed,
in the first quadrant of the complex-$z$ plane.  One must use
Eq.~(\ref{eq:argzp3}), since here $\beta^2 < c$.

While $\Gamma$ is not particularly large, it is much larger than the
$D^*$ width, and the threshold $\sqrt{t_+}$ is much smaller than for
$D_{\ell 3}$ decays since $m_s \ll m_c$.  These effects combine to
give a much larger value of $\gamma$ or $b$.  Table~\ref{tab:Knums}
uses the exact formulae Eqs.~(\ref{eq:zp}), (\ref{eq:abszp}), and
(\ref{eq:argzp3}); the approximations Eqs.~(\ref{eq:abszapprox}),
(\ref{eq:arctanarg}), which drop subleading terms in $\gamma$ or $b$,
give $|z_p| = 0.94366$ ($< 0.2\%$ smaller) and $\arg z_p = 84.0^\circ$
($< 0.2\%$ larger).

Even so, $|z_p|$ does not lie far from the unit circle, and therefore
the typical weakening of the form factor bound $1/|P(0;z_p)| =
1/|z_p|$ as given by Eq.~(\ref{eq:Pexact}) is 1.0578.  Since $z_p$
lies in the first quadrant, from Eq.~(\ref{eq:Blaschke_new}) one finds
that the largest correction to the $a_n$ factors occurs at $z = z_{\rm
max}$ ($t = m_{\ell}^2$), and it equals 1.0581, {\it i.e.}, uniformly
less than 6\%.  Even for the extreme case of $K_{\ell 3}$ decays,
where the above-threshold pole lies far from threshold, the effect on
the parametrization coefficients is quite minimal.

\section{Discussion and Conclusions} \label{sec:Concl}

In this paper we have extended the utility of the model-independent
form factor parametrization for semileptonic decays to explicitly
incorporate the effects of above-threshold resonant poles, such as
$D^{*+}$ in $D^0 \to \pi^- e^+ \nu_e$ and $K^{*+}$ in $K^+ \to \pi^0
e^+ \nu_e$.  Since such poles have a finite width, they lie off the
unitarity cut along the real axis in momentum-transfer space, and
therefore map into the interior of the unit disk in the kinematic
variable $z$.  Inasmuch as the width of such resonances is small
compared to the other mass scales in the system, the pole lies just
inside the unit $z$ circle, and as we showed, consequently has a
rather small effect on the constraints on the form factor
coefficients.

The recipe for calculating the amount of the relaxation of the bounds
due to the presence of an above-threshold pole is easily obtained
through the following steps: First, compute the dimensionless
resonance mass $\mu$ and width $\gamma$ factors directly from $M_R$
and $\Gamma$ using Eqs.~(\ref{eq:mudef})--(\ref{eq:gamdef}), and from
them the dimensionless parameters $a$, $b$, and $c$ using
Eq.~(\ref{eq:polevars}) as well as the dimensionless parameter $\beta$
derived from the threshold $\sqrt{t_+} = M + m$ and adjustable
optimization parameter $t_0$ (or $N$) from Eq.~(\ref{eq:Nopt}).  The
exact position $z_p$ of the pole is then given by
Eqs.~(\ref{eq:zp})--(\ref{eq:argzp1}).  A simple estimate for the
amount of the relaxation of the bounds is given by
Eq.~(\ref{eq:Pexact}), but the full result is obtained by varying the
Blaschke function $1/|P(z;z_0)|$ of Eq.~(\ref{eq:Blaschke_new}) over
the whole allowed semileptonic range for the variable $z$, as given by
Eq.~(\ref{eq:zmaxminopt}).

The Blaschke pole factors present the tremendous benefit of depending
only upon the resonant mass and width, and not upon its residue, a
quantity that is usually much harder to obtain experimentally.  Such a
result is all the more remarkable because models for semileptonic form
factors often assume shapes given by pole dominance, introducing a
source of potentially unquantifiable uncertainties.  If one uses the
techniques in this paper to accommodate above-threshold resonances but
still wishes to obtain tighter bounds on the semileptonic form factors
by incorporating physics along the cut, then only the much milder
multi-hadron continuum dependence of the cut function needs to be
modeled.  Alternately, one may take a minimal (and completely
model-independent) approach by using only the deep-Euclidean
perturbative expression for the relevant two-point function to bound
the form factor integral and hence the allowed parameters defining
each form factor.

\begin{acknowledgments}
  This work was supported by the U.S.\ Department of Energy under
  Grant No.\ DE-SC0009919 (B.G.) and by the National Science
  Foundation under Grant Nos.\ PHY-1068286 and PHY-1403891 (R.F.L.).
  In addition, R.F.L.\ thanks G.~Paz for discussions on the relation
  of poles and resonances.
\end{acknowledgments}


\end{document}